\documentclass[12pt,preprint]{aastex}

\newcommand{\chandra}{{\it Chandra}}
\newcommand{\psr}{PSR J0205+6449}

\slugcomment{Accepted for Publication in {\it The Astrophysical Journal 
Letters}}

\begin{document}

%------------------------------------------------------------------------------
\title{New Constraints on Neutron Star Cooling from {\it Chandra} 
Observations of 3C58}

\author{Patrick Slane\altaffilmark{1}, David J. Helfand\altaffilmark{2},
and Stephen S. Murray\altaffilmark{1}}

\altaffiltext{1}{Harvard-Smithsonian Center for Astrophysics,
    60 Garden Street, Cambridge, MA 02138}
\altaffiltext{2}{Department of Astronomy, Columbia University}
%------------------------------------------------------------------------------

%------------------------------------------------------------------------------
\begin{abstract}
3C58 is a young Crab-like supernova remnant. Historical evidence strongly 
suggests an association of the remnant with supernova SN~1181, which
would make 3C58 younger than the Crab Nebula. Recent \chandra\ observations
have identified the young 65~ms pulsar J0205+6449 at its center, embedded 
in a compact nebula which, we show here, appears to be confined by the 
pulsar wind termination shock. We present new \chandra\ observations of 
this compact nebula and embedded pulsar which set strong upper limits on 
thermal emission originating from the neutron star surface. These limits 
fall far below predictions of standard neutron star cooling, requiring 
the presence of exotic cooling processes in the neutron star core.
\end{abstract}
%------------------------------------------------------------------------------

%------------------------------------------------------------------------------
\keywords{ISM: individual (3C58), pulsars: individual (PSR J0205+6449),
stars: neutron, supernova remnants, X-rays: general}
%------------------------------------------------------------------------------

%------------------------------------------------------------------------------
\section{Introduction}

Neutron stars are macroscopic manifestations of processes that otherwise occur 
only in individual atomic nuclei. Formed hot in the core collapse 
that terminates
the life of a massive star, they are supported against gravitational implosion
by neutron degeneracy pressure. However, details of the interior
structure of neutron stars (NSs) remain poorly understood, largely due to our 
incomplete understanding of the strong interaction at ultrahigh densities. 
In the early stages of their lives, energy loss is dominated by
neutrino emission. However, the neutrino production rate is highly dependent
upon the structure of the interior. In the ``normal'' cooling scenario, 
neutrino production proceeds via the modified URCA process, neutrino
bremsstrahlung in the stellar crust, and plasmon decay.
% the associated neutrino emissivity is $\epsilon_\nu \approx 
%10^{20-21}T_9^8{\rm\ erg\ cm}^{-3}{\rm\ s}^{-1}$ (Page 1997). 
The residual heat diffuses from the core to the surface, manifesting itself as 
blackbody-like emission (modified by effects of any residual atmosphere)
which peaks in the soft X-ray band. The rate at which the surface temperature 
declines depends critically upon the neutrino emission rate, and thus provides
constraints on hadronic physics at high densities.

In some models, the energy density of nucleons in the core
is sufficiently high that the production of pion or kaon condensates, or
quark matter, are predicted. The presence of such exotic particles enhances
the neutrino rate immensely, resulting in much more rapid cooling. 
Consequently, measurements of the surface temperature for NSs of known age
can provide important constraints on NS interiors, and thus
on our knowledge of the strong interaction. Here we present observations
of 3C58, a young Crab-like nebula believed to be associated with the
supernova of 1181 AD.
Given this age, 3C58 is presumably powered by one of the youngest
neutron stars in the Galaxy, and one for which the residual cooling 
emission should be measurable if standard cooling models apply
-- a prospect that has motivated many searches for the pulsar and
its cooling emission. 
The pulsar in 3C58 was recently discovered by Murray et al. (2002). 
To date, however, the strongest upper limit on cooling from the NS surface
yielded a value well in excess of standard cooling curves (Helfand et al.
1995), while Torii et al. (2000) reported hot blackbody emission associated
with a small emitting area that they associated with hot polar caps - a
result inconsistent with {\it XMM} studies by Bocchino et al. (2001).
Here we present {\it Chandra} observations and 
modeling of the emission from the NS which yield a surface temperature
well below that predicted by standard cooling curves, providing strong 
support for the presence of enhanced neutrino production rates in the stellar
interior.

\section{Observations}

3C~58 was observed with the ACIS detector onboard the {\it Chandra} 
Observatory on 04 September
2000. The aimpoint was placed on the S3 chip of the detector, and a
1/2-subarray mode was selected in order to minimize pileup from the compact
central source while still imaging most of the pulsar wind nebula (PWN). 
Standard cleaning of the
data to remove episodes of high background resulted in a final exposure of
33.8~ks. The image clearly reveals a compact central core, as reported by
Murray et al. (2001) 
based on \chandra\ HRC observations, as well as other
apparent structure in the inner nebula -- most notably an arc of emission
apparently emanating from the central region (Figure 1). 

A total of eight relatively bright point sources are observed in the
field defined by the S3 and S4 ACIS chips. Based on identification
of stellar counterparts for four of these, we obtained a position
reconstruction error of $\Delta {\rm RA} = -0.17 \pm 0.12$~arcsec and
$\Delta {\rm Dec} = +0.64 \pm 0.08$~arcsec, within typical accuracies for
\chandra\ positions.

\section{Data Analysis}

\subsection{Spatial Modeling}
The central region of 3C58 is shown in Figure 2. The emission is clearly 
extended, with elongation in the N-S direction, perpendicular to the long
axis of the main nebula. The pulsar \psr\ (Murray et al. 2001) resides 
at the center of this compact nebula. We have carried out spatial modeling of the emission using a
three-component model consisting of a 2-D gaussian combined with a point
source, both superposed on a flat background. For the point source, we used 
a model of the \chandra-ACIS point spread 
function for the mean position of the central region. The free parameters in 
the fit are the central position, widths, rotation angle, and 
normalization of the 2-D gaussian, the position and normalization of the 
PSF model, and the surface brightness of the background. Our best-fit 
parameters for the 2-D gaussian yield a FWHM (along the long axis) of
$4.2^{\prime\prime} \pm 0.04^{\prime\prime}$ and an
ellipticity of $0.49 \pm 0.01$, with the long axis running almost exactly 
north-south. In fact, Figure 2 shows that the full extent
of the elongated core
is $\sim 25^{\prime\prime}$ in the N-S direction.
Murray et al. (2002) find similar results using HRC data, though a somewhat 
smaller extent which may be the result of the factor of $\sim 2.5$ fewer 
counts in the HRC image, as well as the fact that the 2-D gaussian model is
only a rough approximation of the more extended structure.
The best-fit point source location, when corrected for the offsets
described above, is RA$_{2000}$: 02$^h$05$^m$37.92$^s$,
Dec$_{2000}$: $+64^\circ$49$^\prime$42.8$^{\prime\prime}$; the total
number of counts from this component is $880 \pm 78$ (90\% confidence).

The western edge of the core nebula lies directly along a radio 
filament (Frail \& Moffett 1993), shown as contours in Figure 2,
and there appears to be a slight flattening of nebula along this
side. Frail \& Moffett (1993) suggested that this filament may represent
the position of the termination shock where the pulsar wind is confined
by the interior pressure of the PWN. Adopting the generally accepted distance $d=3.2d_{3.2}$~kpc (see Section 4.1), integration of the radio 
synchrotron spectrum yields a pressure of 
$P_{\rm\ neb} = 3.2 \times 10^{-10} d_{3.2}^{-1}{\rm\ dyne\ cm}^{-2}$
under the assumption of equipartition between the electron and magnetic energy
densities. The ram pressure of the pulsar wind
$P_{\rm\ wind} = \dot E/4 \pi \eta c r_w^2$, 
where the wind covers a fraction $\eta$ of a sphere, $\dot E$ is the
energy loss rate of the pulsar, and $r_w$ is the distance from the pulsar
at which wind confinement occurs. The spin-down properties of the pulsar
give $\dot E = 2.6 \times 10^{37} I_{45} {\rm\ erg\ s}^{-1}$ (Murray
et al. 2001) where $I_{45}$ is the NS moment of inertia
in units of $10^{45}{\rm\ gm\ cm}^{2}$. Pressure balance should thus
occur at 
$r_w = 5.5 \times 10^{17} I_{45}^{1/2} \eta^{-1/2} d_{3.2}^{1/2}$~cm, 
or at an angular distance 
$\theta = 11.4 I_{45}^{1/2} \eta^{-1/2} d_{3.2}^{-1/2}$~arcsec. 
This is in good agreement with the the $\sim 12$~arcsec radial extent of the 
core X-ray emission in the NS direction, suggesting that the compact nebula 
surrounding the pulsar is bounded by the pulsar wind termination shock. It is 
possible that this is actually a toroidal structure, much like 
that seen in the Crab Nebula, and that the elongated surface brightness
distribution is the result of
the inclination angle. In this interpretation, the axis of the toroid,
which presumably lies along the rotation axis of the pulsar, lies in the 
east-west direction when projected onto the sky. We note that the long
axis of 3C58 itself, as well as an extended jet-like feature shown in Figure 2,
are both aligned in this direction.
Assuming that the radio filament lies along one side of the torus, its 
separation from the pulsar ($\sim 4.5$~arcsec) implies an inclination
angle of $\sim 70^\circ$. We defer detailed discussion of this picture,
including the jet-like feature in Figure 2 and other possible outflow-related
structures seen in Figure 1, to a future publication. 

\subsection{Spectral Modeling}

Previous measurements of the column density for 3C58 yield 
$N_H = (3.3 \pm 0.4) \times 10^{21}{\rm\ cm}^{-2}$ (Torii et al. 2000).
This is consistent with values reported by Bocchino et al. (2001).
Using the entire nebula, and allowing the photon index to vary with 
radius\footnote{A variation in spectral index with radius is expected from 
synchrotron cooling.  These results will be presented in a future publication.}
we obtain $N_H = (3.75 \pm 0.11) \times 10^{21}{\rm\ cm}^{-2}$. Using
this value, 
the spectrum from the central $3 \times 3$ pixel region\footnote
{Each ACIS pixel is 0.492~arcsec on a side.} in 3C58 is well fit 
($\chi^2_r = 0.82$) by an absorbed power
law with a photon index $\Gamma = 1.73 \pm 0.07$ (Figure 3); blackbody
or thermal plasma models are ruled out at high significance. 
In particular, contrary to results from ASCA studies (Torii et al. 2000), we 
find no evidence of a soft blackbody component from the central source,
which is consistent with the results of Bocchino et al. (2001).
The majority of equations of state yield an effective
NS radius larger than 12~km for any range of masses (Haensel 2001).
To establish an upper limit to the effective blackbody temperature of any
such a component, we have fixed the column density at the
upper end of the 90\% confidence interval quoted above, added a blackbody
model with $R_\infty = 12$~km to the power law, and varied
the temperature of this component  until the spectral fit probability fell 
to $10^{-4}$. In re-fitting the data, we allowed the power law index to
vary as well.
%\footnote{Fixing the power law index at the best-fit value
%listed above yields an even stronger constraint on $T.$}
The upper limit we determine is $T_\infty < 1.08 \times 10^{6}$~K. The
model containing this thermal component is shown in Figure 3, and clearly
exceeds what can be accommodated in the data. For comparison, we also plot the
expected flux for a blackbody component whose temperature is that expected from
standard neutron star cooling curves.

As with all stars, the emission from the surface of a NS is not a blackbody; 
rather, it is modified by the presence of whatever atmosphere might exist. 
One expects the surface of the NS to be covered with Fe, but an atmosphere
consisting of H, He, and/or intermediate-mass elements acquired either from
ejecta fallback following 
the neutron star's formation, or from material accreted from the ISM, is also a
possibility. From models of nonmagnetic atmospheres, the primary 
effect of H or He atmospheres is a considerable deviation of the 
high energy end of the spectrum relative to the Wien tail of a pure
blackbody. The result is that the best-fit blackbody model 
overestimates the effective temperature -- typically by as much as a factor
of two. For atmospheres dominated by heavier elements the blackbody fit
gives a good approximation to the temperature (Lloyd, Hernquist, \& Heyl 2002).
The temperature upper limit derived above is thus conservative. 

\subsection{Spatial/Temporal Modeling}

The temporal information from \psr\ provides an additional constraint
on any thermal emission originating at the surface of the NS.
The spectral results summarized above clearly establish that the bulk
of the flux from the pulsar is nonthermal. The narrow pulse profile measured
by Murray et al. (2001) is consistent with this picture. While temperature
nonuniformities on the NS surface can, in principle, lead to a pulsed
X-ray signal, light bending effects from the strong gravitational field
result in considerable broadening of any such pulse profile (Yancopoulos,
Hamilton \& Helfand 1994; Zavlin, Shibanov \&  Pavlov 1995)
The very sharp pulse profile of \psr\ is inconsistent with such
an origin. We can thus use the strength of the pulsed signal along with 
the modeled brightness of the pointlike component to obtain an independent
limit on any remaining emission arising from surface cooling. 

In the 33~ks HRC-S observation of 3C58,
the pulse profile for \psr\ yields a total of $\sim 127.2 \pm 17.2$ counts 
(90\% confidence) above
the mean flux level in the narrow phase bins containing the bulk of the 
pulsed power. 
The count rate conversion between ACIS-S and HRC-S based
on the measured power law spectrum reported above is 
$R_{\rm HRC-S} = 0.26 R_{\rm ACIS-S}$.
The lower limit to the number of pulsed counts in the $\sim 34$~ks ACIS-S 
observation is thus $\sim 435$.
Our spatial modeling yields a total of $880 \pm 78$ counts from the point
source in ACIS-S. The 90\% confidence upper limit to the number of 
unpulsed counts is
thus $\sim 523$, for an ACIS-S count rate of $1.54 \times 10^{-2}{\rm\ ct\
s}^{-1}$. For a radius at infinity of 12~km, the upper limit to the
blackbody temperature is $1.13 \times 10^6$~K, consistent with the limit
obtained through spectral modeling. As we discuss below (and show in Figure
4), this limit falls well below values predicted for standard cooling models
assuming the historical age of 3C58. We note that the limit depends on 
uncertainties in the distance and column density. However, even at the most
extreme values of acceptable column density based on fits to the entire nebula,
the distance required to match the standard cooling predictions is $\sim
6$~kpc, a value inconsistent with that inferred from H\,{\sc i} 
measurements.

\section{Discussion}

Our measurement of an upper limit to the temperature of the young
NS in 3C58 provides the strongest constraint on NS
cooling yet derived, and poses significant problems for standard cooling
models. We explore here the implications of our result
for the nuclear equation of state and the structure and evolution of
NS.

\subsection{Neutron Star Cooling}

The cooling rate of isolated NSs has been a subject of
considerable theoretical work predating even the discovery of the first
pulsars (e.g. Bahcall \& Wolf 1965a,b). The poorly understood properties of
the strong nuclear potential at the densities found in NS interiors 
make these calculations difficult, and lead to a wide range of predictions
based on different assumptions for the equation of state, composition, and
details of superconductivity (see, e.g., the review by
Tsuruta 1998, and references therein). While there is a clear consensus
that the early cooling proceeds via neutrino emission from the NS core, the
time scale over which this dominates depends critically on the neutrino
production rate which, in turn, can vary by orders of magnitude depending upon
the state of matter in the interior.

Broadly speaking, models can be divided into ``standard'' and ``non-standard''
cooling scenarios. The former class is based upon neutrino emission via
the modified URCA process, n--n and p--p neutrino bremsstrahlung, crust
neutrino bremsstrahlung, and plasmon neutrino processes. Assumptions must
still be made in the cooling calculations, such as 
the form of the the nucleon-nucleon force and the many-body technique
used to determine the equation of state, but the differences between stiff
and soft equations of state are not large for this class of models. In Figure
4, we plot a variety of cooling curves associated with different models for
the NS interior and its properties (see Page 1998 and references 
therein)\footnote{Tabulated values are available at
http://www.astroscu.unam.mx/neutrones/NS-Cooler/NS-Cooler.html}.
The solid curve corresponds to ``standard''
cooling using an equation of state of moderate stiffness. 

Non-standard cooling models incorporate neutrino emissivities 
associated with other processes that may operate in NS interiors. These 
include the presence of pion condensates which may form 
%via $n \rightarrow p + \pi^-$ 
at sufficiently high densities. The resulting pion-induced
beta decay leads to very a high neutrino emissivity and a correspondingly
shorter cooling time for the NS interior (Bahcall \& Wolf 1965a,b, Maxwell et 
al.  1977). Similar processes involving kaon
condensates or quark matter may operate as well. Alternatively, strong magnetic
field effects in the crust may allow the normal URCA process to proceed, which
also leads to extremely high neutrino production rates. These exotic cooling
mechanisms modify the NS cooling curves substantially. The dashed curves
in Figure 4 represent approximations for several nonstandard cooling models 
to illustrate the associated rapid cooling (Page 1998). The effects of 
superfluidity can substantially 
moderate the rapid cooling because the significantly reduced heat
capacity of the superfluid particles reduces the neutrino rate 
considerably.

We have indicated on Figure 4 the upper limit for surface cooling emission
from \psr\ derived above. It is evident that the limit falls considerably below
predictions from standard cooling models, suggesting the presence of some
exotic cooling contribution in the interior. We note that the placement of
this point on the cooling curves depends upon the age assumption for 3C58.
Here we have used the historical age based on association with SN~1181.
Four independent written records -- two from Japan, and one each from the
northern and southern Chinese empires -- record the presence of a guest star
that appeared in the late summer of AD 1181 and was visible to the naked eye for
nearly 6 months. Detailed analysis of the star's celestial
coordinates are summarized by Stephenson and Green (1999) who identify a
faint star near $\epsilon$ Cas as the closest permanent star to the location
of the event. This position lies less than $2^{\circ}$ from 3C58. The only 
other young supernova remnant within $10^{\circ}$ is Tycho's SNR, the 
unambiguously identified remnant of SN~1572. The association of 3C58 with
SN~1181 thus appears extremely secure.
However, even if we increase the age estimate to 5000~yr, as has been
suggested based on optical and radio expansion data (Fesen 1985,
Bietenholz et al. 2001), the measurement still falls below standard 
cooling models (see Figure 4). Deeper
high-resolution X-ray measurements of 3C58 are of considerable interest
in order to further constrain this important diagnostic of the NS
interior.

\section{Summary}

The size of the extended emission region immediately surrounding the pulsar 
in 3C58 is consistent with that expected for the termination shock where the pulsar
wind pressure is balanced by the interior pressure of the nebula. The elongated
morphology, along with the position and size of an apparently associated radio
filament, suggest a toroidal morphology similar to that observed in the Crab
Nebula. The orientation of the torus suggests a pulsar rotation axis
aligned in the east-west direction, along the primary axis of extension of
3C58 itself. The inclination angle of the torus is large, and the resulting
projected X-ray morphology does not permit us to resolve the toroidal structure
with the current X-ray data.

Our \chandra\ observations of 3C58 and its associated pulsar provide
important constraints on models for NS cooling. The upper limit
to the thermal emission from the NS surface falls well below
predictions from standard cooling models and appears to require the
presence of some rapid cooling mechanism.
Additional time-resolved observations of the pulsar could further constrain
this limit, providing important new information on the equation of state
of nuclear matter and the structure of NS interiors.

%------------------------------------------------------------------------------

%------------------------------------------------------------------------------
\acknowledgments
The authors would like to thank John Bahcall for helpful discussions on 
NS cooling and Dale Frail for supplying the VLA image of the 
central region in 3C58. We also thank Bryan Gaensler for his helpful
suggestions and careful reading of the text.
This work was supported in part by the National Aeronautics and
Space Administration through contract NAS8-39073 and grant GO0-1117A (POS)
and contract NAS8-38248 (SSM).

%------------------------------------------------------------------------------

%\begin{thebibliography}{}

%\end{thebibliography}

%------------------------------------------------------------------------------

\clearpage


\begin{figure}
\plotone{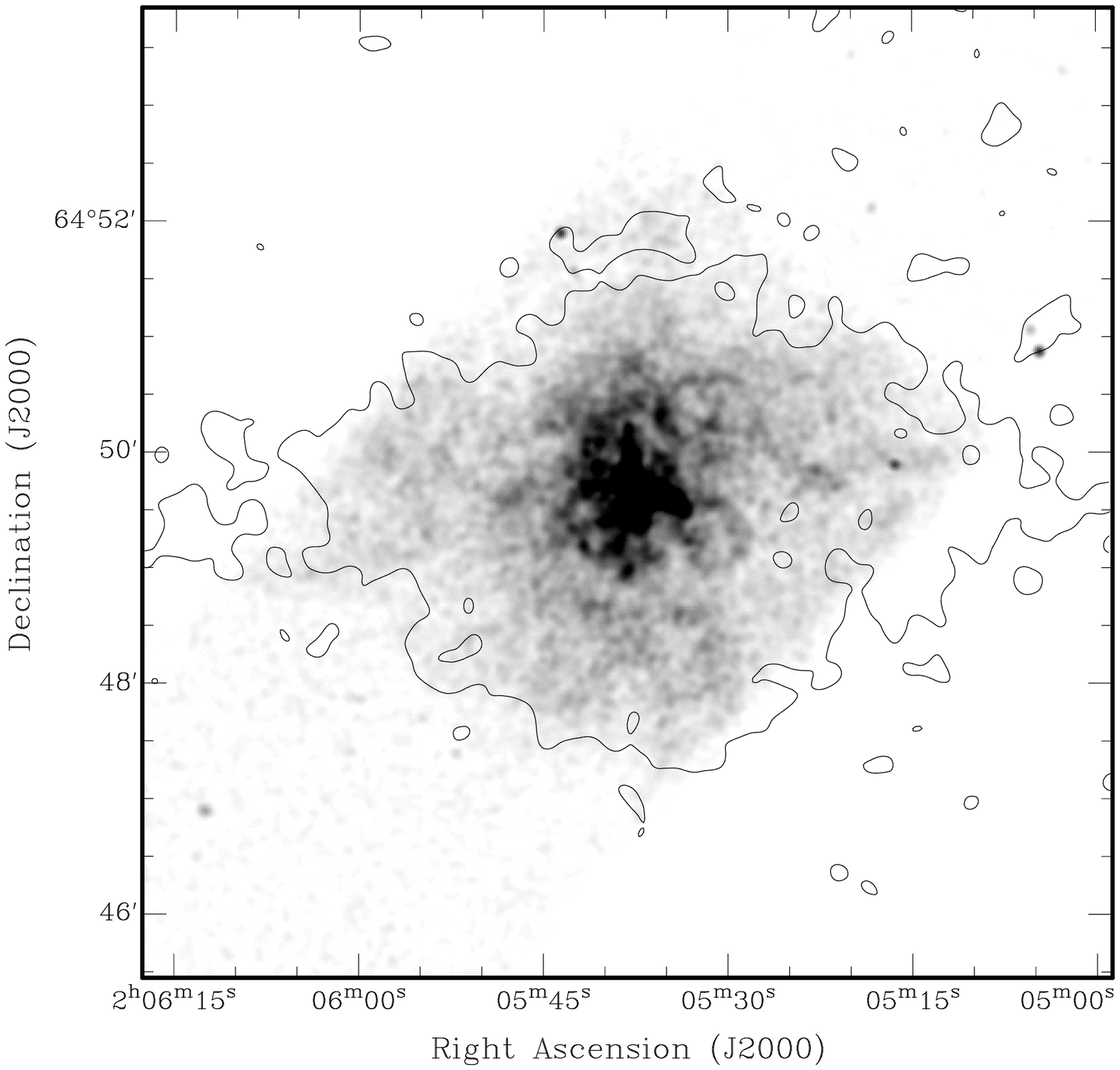}
\caption{ACIS image of 3C58. Some portions of the nebula extend
beyond the detector boundary, as indicated by the outermost contour from the
HRC image, which is superimposed.}
\end{figure}

\clearpage

\begin{figure}
\plotone{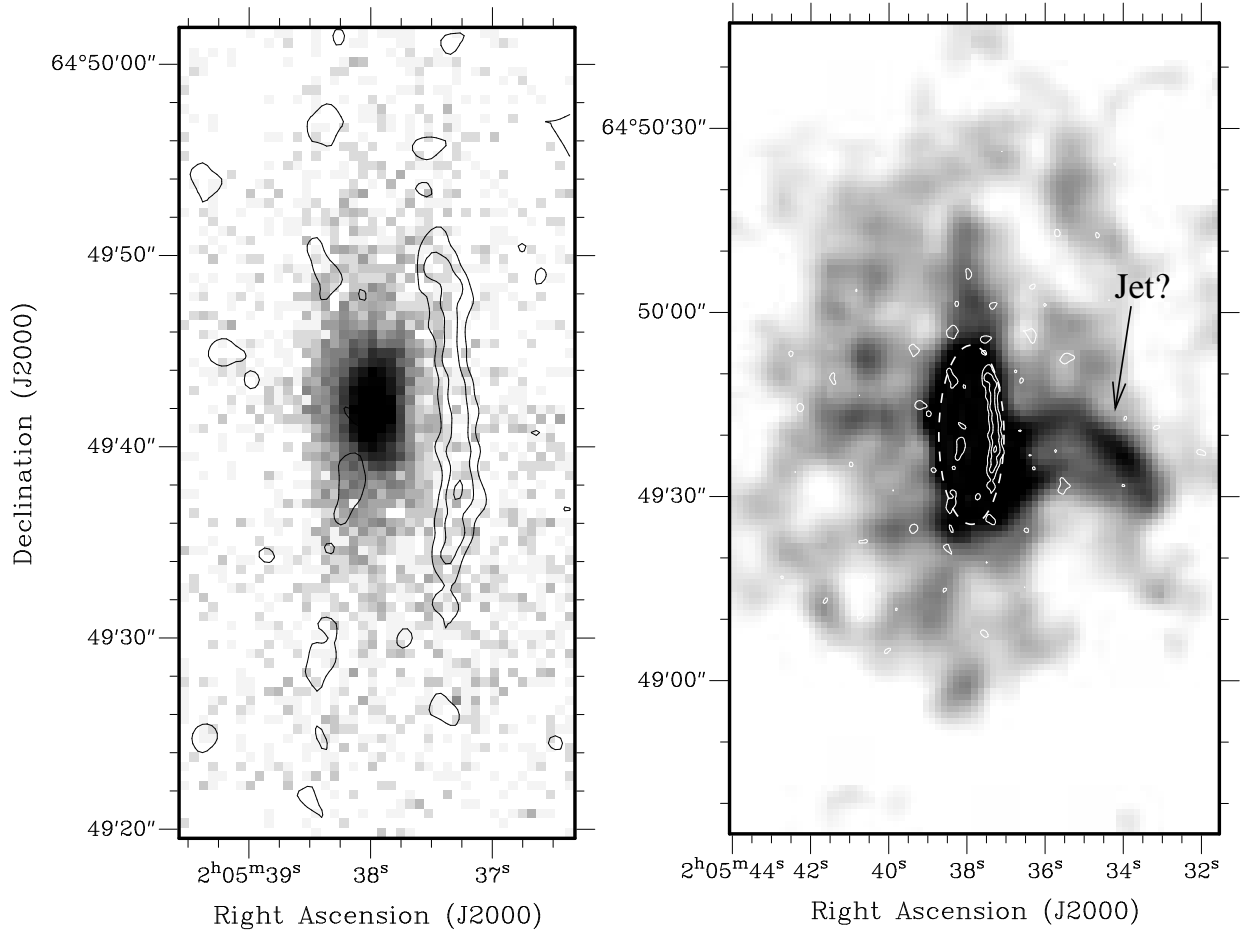}
\caption{{\bf Left:} 
ACIS image of the central region of 3C58, with contours from 20~cm VLA data 
showing a faint radio wisp that bounds the X-ray core.
{\bf Right:} Saturated image of 3C58 core, on a larger scale, revealing
the faint jet-like feature extending toward the west. The dashed ellipse
indicates the rough outline of the extended X-ray core.}
\end{figure}

\clearpage

\begin{figure}
\plotone{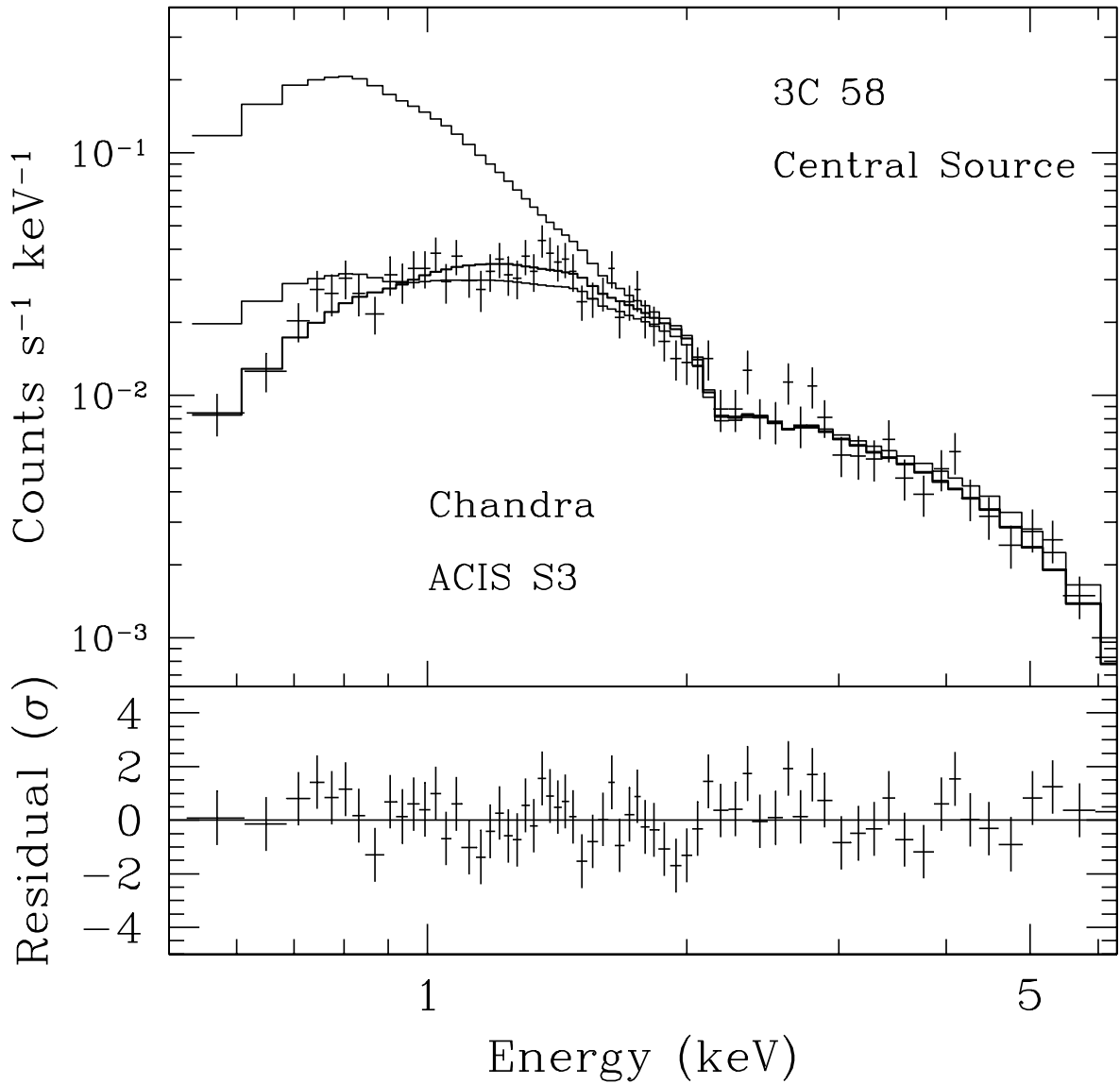}
\caption{ACIS-S spectrum of the central $3 \times 3$ pixel region 
($1.5^{\prime\prime} \times 1.5^{\prime\prime}$) centered on \psr. The 
lower histogram (matching the data) corresponds to the best-fit 
power law model. There is no 
evidence for a soft blackbody-like component. The uppermost histogram depicts
a model with an additional component representing blackbody emission from 
a NS with effective radius 12~km, at a temperature consistent with standard
cooling models. The middle histogram (at low energies) corresponds to the same
model, but using a temperature determined as the upper limit that could have 
been present in the 3C58 data.
}
\end{figure}

\clearpage

\begin{figure}
\plotone{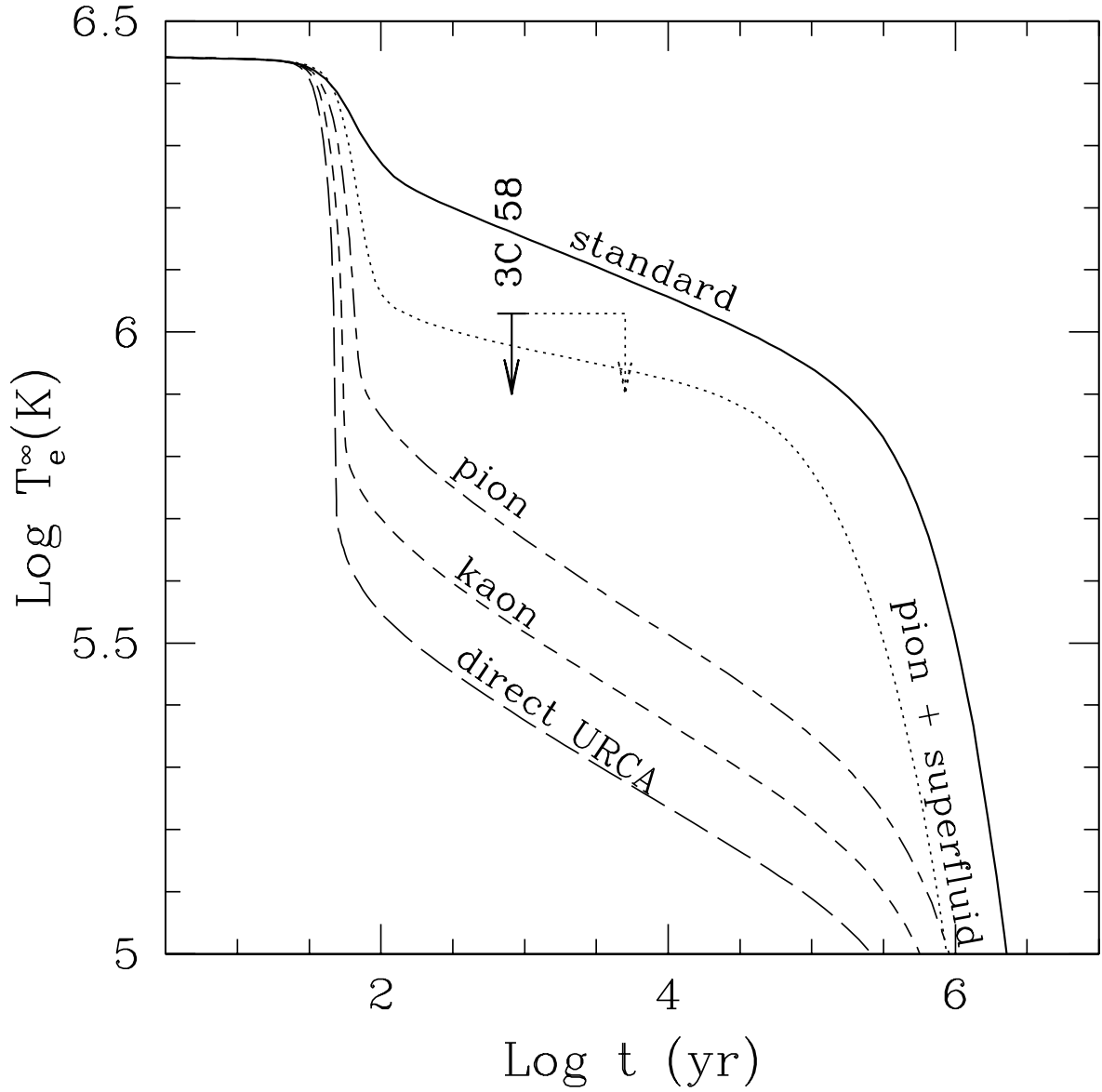}
\caption{
Surface temperature upper limit for 3C58, in comparison with
predictions for cooling of NS. Model data are from Page 1998. 
The dashed arrow for 3C58 illustrates the upper limit for the pulsar
if the characteristic age, rather than the historical age, is used.
}
\end{figure}

\end{document}